\documentclass[aps,prd,twocolumn,superscriptaddress,nofootinbib]{revtex4}

\usepackage{amsmath,amssymb,graphicx}
\usepackage{soul}

\usepackage{color}

\newcommand{\be}{\begin{eqnarray}}
\newcommand{\en}{\end{eqnarray}}
\newcommand{\nn}{\nonumber\\}
\newcommand{\ctk}{\chi(t,\mathbf{k})}

\newcommand{\chir}{\chi_{\rm{R}}}
\newcommand{\phic}{\phi_{\rm{crit}}}
\newcommand{\sigmac}{\sigma_{\rm{crit}}}
\newcommand{\fnl}{f_{\rm{NL}}}
\newcommand{\mpl}{m_{\rm{pl}}}
\newcommand{\ns}{n_{\rm{s}}}

\begin{document}
\title{Supersymmetric hybrid inflation with a light scalar}

\newcommand{\addressImperial}{Theoretical Physics, Blackett Laboratory, Imperial College, London, SW7 2AZ, United Kingdom}

\author{Stefano Orani}
\affiliation{\addressImperial}
\affiliation{Deparment of Physics, University of Basel, Klingelbergstrasse 82, CH-4056 Basel, Switzerland}

\author{Arttu Rajantie}
\affiliation{\addressImperial}

\date{\today}

\begin{abstract}
Light scalar fields present during inflation can lead to interesting observable signatures, especially in models with non-equilibrium reheating dynamics. We study a supersymmetric hybrid inflation model with a third scalar whose lightness is protected by symmetry, using analytical and numerical techniques, and demonstrate that the amplitude, the spectral index and the non-Gaussianity parameter $\fnl$ of the primordial curvature perturbation are within the Planck observational bounds for suitable parameter values.

\end{abstract}

\maketitle

\section{Introduction}

The recently announced measurements of the cosmic microwave background (CMB) anisotropies by the Planck satellite~\cite{Ade:2013zuv,Ade:2013ydc} provided even more evidence for inflation, a period of accelerating expansion in the early universe. This acceleration is believed to have been caused by a scalar field called inflaton, but so far observations have shed little light on the detailed properties of this field. 

In particular, the inflaton should ultimately be incorporated as part of the same quantum field theory that describes particle physics at high energies beyond the Standard Model. One of the most promising attempts to achieve this is hybrid inflation~\cite{Linde:1993cn} 
embedded in supersymmetric (SUSY) theories such as F-term versions in $N=1$ SUSY, D-term versions in supergravity, P-term versions in $N=2$ SUSY, D-brane versions and many more (see~\cite{minSUSY,SUSY} for a few representative examples).

Recent work has also demonstrated that inflationary models with light scalar fields, besides the inflaton field, can have interesting and distinctive observable signatures.
The simplest example is the generation of curvature perturbations through the curvaton mechanism~\cite{curv}. If the light scalar is coupled to the inflaton and reheating at the end of inflation involves non-equilibrium processes the observational signatures are richer, including anisotropies in the gravitational wave background~\cite{Bethke:2013aba} and highly non-Gaussian contributions to the curvature perturbation~\cite{preh}. If the perturbations are subdominant, they would generally lead to a small non-Gaussianity parameter $f_{\rm NL}$ on large scales~\cite{Suyama:2013dqa} and would therefore be compatible with the Planck data.

The effects of non-equilibrium dynamics of light scalars have been studied mostly in the context of preheating with a parametric resonance, but analogous observable signatures should also be produced during tachyonic preheating~\cite{tacpreh1,tacpreh2} in hybrid inflation models. Depending on the details of the theory, tachyonic preheating can involve highly non-trivial non-equilibrium phenomena such as  formation of topological defects, Q-balls or oscillons, which can all be influenced by the light scalar field and therefore lead to observable signatures.

The simplest hybrid inflation model consists of two fields, the inflaton $\phi$ and the waterfall field $\chi$. When inflation ends rapidly, which is required for non-equilibrium processes, the waterfall field has to be heavy, and therefore it would have no effect on cosmological scales.
In bosonic models it is possible to add another scalar field by hand, but if the new field is coupled to the two other fields, it is generally not light (previous works on this topic include \cite{Bernardeau:2004zz,Bernardeau:2002jf,Salem:2005nd,Alabidi:2006wa,Sasaki:2008uc,Naruko:2008sq,Lyth:2005qk}). It is therefore interesting to consider the scenario in the context of SUSY hybrid inflation in which case the lightness of the field is protected by supersymmetry. 

The aim of this paper is to investigate a simple modification of the standard F-term SUSY hybrid inflation model which has a light scalar field. We analyze an inflationary scenario in which the waterfall field is heavy and the symmetry breaking phase transition is fast, in accordance with standard hybrid inflation. We study the primordial perturbations generated on super-horizon scales and find that the model can produce the observed amplitude and spectral index.

\section{The Model}
\label{sec:SUSYhi}

 Hybrid inflation models derived from SUSY can be of two types, F-term and D-term hybrid inflation (for a review of inflationary model building from SUSY see~\cite{SUSYinf}). Of the two, the F-term type attracts more attention because it naturally fits with the Higgs mechanism. In its simplest form, F-term SUSY hybrid inflation is given by the superpotential~\cite{minSUSY}
\be 
W=\alpha\Phi\left(X\overline{X}-\frac{v^2}{2}\right),
\label{susyhi}
\en
 where $\Phi$ is a gauge singlet containing the inflaton and $X$, $\overline{X}$ is a conjugate pair of superfields transforming as non-trivial representations of a gauge group. It is the most general form of superpotential consistent with the R-symmetry, under which $W\rightarrow e^{i\gamma}W$, $\Phi\rightarrow e^{i\gamma}\Phi$ and $X\overline{X}$ is invariant.

 The scalar potential is given by ~\cite{SUSYinf}
\be
V=\sum_\alpha\left|\frac{\partial W(S^{\alpha})}{\partial S^{\alpha}}\right|^2,
\label{susyvdef}
\en
where the sum is taken over all the superfields $S^{\alpha}$.
 For the superpotential (\ref{susyhi}), this gives
\be
V=2\alpha^2|\Phi|^2|X|^2+\alpha^2\left|X\overline{X} - \frac{v^2}{2}\right|^2.
\label{Vhi}
\en
It has a single global minimum,
\be 
X =\pm \frac{v}{\sqrt{2}}, \;\;\;\;\;\;\;\;\; \Phi=0,
\en
which preserves SUSY.

At tree level, there is no term in Eq.~(\ref{Vhi}) that would drive $\Phi$ to zero. It is no longer the case when radiative corrections are taken into account. For $|\Phi|>0$ SUSY is broken, meaning that one-loop corrections to the potential do not cancel. They are given by~\cite{minSUSY}
\be 
\Delta V = \sum_{i}\frac{(-1)^F}{64\pi^2}M_i^4\ln\frac{M_i^2}{\Lambda^2},
\label{1loop}
\en
where $F=0,1$ respectively for a boson or a fermion and $M_i$ is the effective mass measured at a scale $\Lambda$. These corrections lift the potential in the $\Phi$ direction, driving the system towards the global minimun.

 This model is appealing for a series of reasons, perhaps the main one being the possibility to embed it in the particle physics framework. Furthermore, the potential (\ref{Vhi}) has only two parameters, $\alpha$ and $v$. 

The dynamics of the phase transition depend on the hierarchy between the masses of the two fields $\Phi$ and $X$, with behaviour interpolating from a fast phase transition that happens in less than one e-fold to transitions such that considerable expansion happens after the critical point, predicting a red spectral index~\cite{Clesse:2012dw}.  An alternative regime consists of an inflationary phase dominated by the mass term of $\Phi$~\cite{Mulryne:2011ni}. This scenario is unlikely to be realised in the context of SUSY, since $\Phi$'s lightness is protected by supersymmetry. However, considered in this regime, hybrid inflation generates a red spectral index and can lead to a viable cosmology.

Let us now modify the superpotential to introduce a third scalar field which is light at horizon crossing.
The superpotential is
\be W=\alpha\Phi\left(X\overline{X}-\frac{v^2}{2}\right)+\frac{m}{2}\Phi^2, \label{w}\en
where $\Phi$ is a gauge singlet. 
It is convenient to parameterise the extra term using the dimensionless combination $c\equiv m/\alpha v$.

The Kahler potential is supposed to give the canonical kinetic terms plus negligible terms. Note that in Eq.~(\ref{w}) $\Phi$ breaks the R-symmetry. 
This is not a problem because no observation tells us that this is a fundamental symmetry of nature, and since the symmetry is restored as $c\rightarrow 0$,  it is natural to have $c\ll 1$.

The scalar sector of the potential is given by
\be V &=& \alpha^2\left(X\overline{X}-\frac{v^2}{2}\right)^2 + m^2|\Phi|^2 +2\alpha^2|\Phi|^2|X|^2\nn 
         &&  + \alpha m (\Phi + \Phi^*)\left(X\overline{X}-\frac{v^2}{2}\right)\,. 
\label{vscalar}
\en
Let us expand the field $\Phi$ into its real and imaginary components:
$\Phi = \frac{1}{\sqrt{2}}(\phi + i\sigma)$.
The potential (\ref{vscalar}) becomes
\be 
V &=& \frac{\alpha^2}{4}\left(\chi^2 - v^2\right)^2 + \frac{m^2}{2}(\phi^2+\sigma^2)\nn
     && +  \frac{\alpha^2}{2}\chi^2(\phi^2+\sigma^2) + \frac{\alpha m}{\sqrt{2}}\phi\left(\chi^2-v^2\right)\,,
 \label{vs}
\en
where, without loss of generality, we constrained $X=\overline{X}=\chi/\sqrt{2}$, $\chi$ being a real scalar field. Note that the fields $\phi$ and $\sigma$ share the same mass, whereas $\chi$ does not.

The potential (\ref{vscalar}) has three degenerate global minima with $V=0$:
a symmetry-preserving vacuum (SPV) at 
\begin{equation}
\phi = w, \quad \chi=\sigma=0,
\end{equation}
with $w=\alpha v^2/(\sqrt{2}m)$, and
two symmetry-breaking vacua (SBV) at 
\begin{equation}
\phi=\sigma=0,\quad \chi=\pm v.
\end{equation}

We are interested in inflationary trajectories that end in the SBV, with a fast phase transition driven by the waterfall field $\chi$. Its mass is given by $m^2_{\chir} = -\alpha^2v^2+\alpha^2(\phi^2+\sigma^2)+\sqrt{2}\alpha m\phi$ and it becomes tachyonic inside the circle in the $\phi-\sigma$ plane defined by
\be 
\left(\phi+\frac{m}{\sqrt{2}\alpha}\right)^2+\sigma^2=v^2+\frac{m^2}{2\alpha^2}\,.
\label{circle}
\en

\begin{figure}
\centering
\includegraphics[width= 8cm]{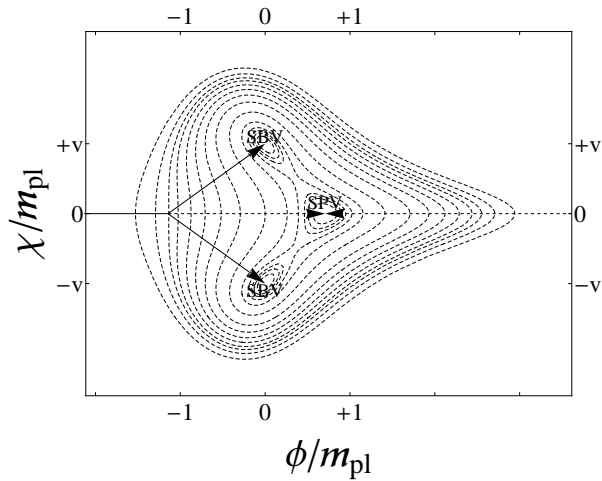}
\caption{Contour plot of the potential $V(\phi,\chi,0)$. We choose $m=v=\mpl$ and $\alpha=1$ for drawing purposes, although these parameters are unrealistic. The arrows represent possible inflationary trajectories. If inflation starts at some $\phi_{\rm i} > 0$, then the trajectory hits the SPV. On the other hand, if $\phi_{\rm i} < 0$, the trajectory may end in the SBV.} 
\label{Vphichi}
\end{figure}

\begin{figure}
\centering
\includegraphics[width= 8cm]{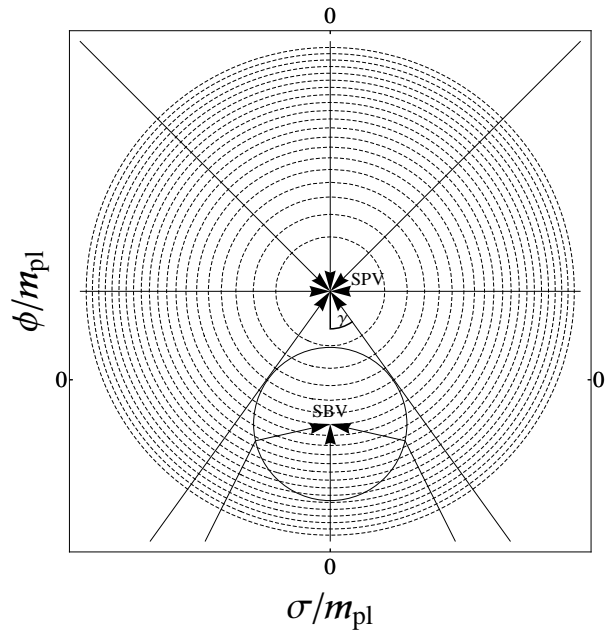}
\caption{Contour plot of the potential $V(\phi,0,\sigma)$. We choose $m=v=\mpl$ and $\alpha=1$ for drawing purposes, although these parameters are unrealistic. The arrows represent possible inflationary trajectories and the circle corresponds to $m^2_\chi = 0$. Trajectories at an angle greater than $\gamma$ to the $\sigma=0$ line don't hit the $m^2_\chi<0$ region and the SBV.} 
\label{Vphisigma}
\end{figure}

The possible inflationary trajectories are shown in Figs.~\ref{Vphichi} and \ref{Vphisigma}. The key fact is that, at large field values, the global attractor of the system is the SPV. Only trajectories that, on their way to the SPV, hit the tachyonic region $m^2_\chi<0$, can potentially reach the SBV, as can be seen in Fig.~\ref{Vphichi}. However, hitting the instability region is not sufficient to guarantee that the system ends up in the SBV: the waterfall field $\chi$ still has to roll sufficiently far down the ridge while in the tachyonic region. Otherwise the trajectory simply crosses the instability region and ends up in the SPV. This means the waterfall field has to be heavy: $\alpha v \gg m$ or, equivalently $c \ll 1$. For such parameters, the waterfall is fast, happening in much less than one e-fold.

It is easy to see from Fig.~\ref{Vphichi} that trajectories rolling from positive $\phi$ end up in the SPV. This constrains the interesting trajectories to those with negative initial values, 
$\phi<0$. 
Fig.~\ref{Vphisigma} strengthens the constraint: Only trajectories within an angle $\gamma$ from the $\sigma=0$ trajectory in the $\phi-\sigma$ plane hit the instability region. Assuming that the probability distribution of the initial conditions along a circle centred at the SPV of large radius in the $\phi-\sigma$ plane is uniform, the likelihood of a trajectory hitting the instability region is given by
\begin{equation}
\label{gamma}
P_{\rm SBV}=\frac{\gamma}{\pi}=\frac{1}{\pi}\arcsin\frac{c\sqrt{c^2+2}}{c^2+1}
\approx\frac{\sqrt{2}}{\pi}c. 
\end{equation}
In order to avoid fine-tuning of initial conditions, $P_{\rm SBV}$ should not be too small, and therefore we will focus on the largest possible ratio $c$ that leads to the desired phenomenology. As we shall see, this ratio turns out to be $c\approx10^{-3}$, which gives a likelihood $P_{\rm SBV} \approx 10^{-3}$, and therefore fine tuning cannot be completely avoided. Of course, as usual, it may be possible to justify the fine tuning by anthropic reasoning.

\section{Perturbations} 
\label{sec:deltaN}

To compare the model with observations, we compute the spectral index, the magnitude of $\fnl$ and the amplitude $A_\zeta$  of the power spectrum using the $\delta N$ formalism~\cite{Starobinsky:1986fxa, Sasaki:1995aw, Lyth:2005fi}, implemented with numerical and analytic techniques.

The $\delta N$ formalism relies upon the separate universe approximation, according to which points in the 
universe that are causally disconnected evolve like a 'separate universe', with dynamics following from the local energy density and pressure, 
and obeying the field equations of a Friedmann-Robertson-Walker (FRW) universe (see for example~\cite{Lyth, Wands:2000dp}). 
The $\delta N$ formalism then notes that the uniform density curvature perturbation between two such points, $\zeta$, is given by the difference in 
e-folds of expansion between them, from some initial 
shared flat hypersurface labelled $*$, to some final 
shared hypersurface of uniform density labelled $f$,
\be
\zeta = \delta N^f_*.
\en 
We use $*$ to label the flat hypersurface, since 
in order to make observational predictions for an inflationary 
model, this surface must be 
defined at the time when observable scales exited the cosmological 
horizon. Moreover, 
the final hypersurface $f$ must be taken at some much 
later time when the dynamics are adiabatic
and $\zeta$ is conserved.
This can happen, for example, long after inflation ends, 
when the dynamics are 
dominated by a single fluid.

When slow-roll is a good approximation at horizon exit, $\zeta$ 
is completely determined by
the field perturbations $\delta\varphi^\alpha_*$ at that time,
\be
\zeta=\delta N_*^f(\delta\varphi^\alpha_*).
\en
The field perturbations are extremely close to 
Gaussian~\cite{Gaussian}, and if their amplitude 
is sufficiently small, the statistics of 
the curvature perturbation can be determined in a 
simple manner by Taylor expanding
\be
\delta N^f_* \approx \frac{\partial N^f_*}{\partial \varphi^{\alpha}_*} \delta\varphi^{\alpha}_* + \frac{1}{2}\frac{\partial^2 N^f_*}{\partial \varphi^{\alpha}_*\varphi^{\beta}_*} \delta\varphi^{\alpha}_*\delta\varphi^{\beta}_*\,,
\label{eq2}
\en
where here and from here on we employ the summation convention, and subsequently we 
will employ the notation in common use, $N_{\alpha}={\partial N^f_*}/{\partial \varphi^{\alpha}_*}$.

One then finds that the amplitude of the power spectrum
is given by 
\be
A_\zeta^2 = N_{\alpha}N_{\alpha}\frac{H_*^2}{4\pi^2}\,,
\label{eq:ampl}
\en 
and the spectral index $\ns$  
by~\cite{Sasaki:1995aw,largeNG1b} 
\be
\ns =1   - 2\epsilon_* + \frac{2}{H_*}\frac{\dot{\varphi}^{\alpha}_*N_{\beta}N_{\alpha\beta}}{N_{\alpha}N_{\alpha}} \,.
\label{eq:spectrum}
\en
The non-Gaussianity of the perturbation is characterised by the reduced bispectrum $\fnl$, which is given by~\cite{Lyth:2005fi}
\be
 \fnl = \frac{5}{6}\frac{N_{\alpha}N_{\beta}N_{\alpha\beta}}{(N_{\alpha}N_{\alpha})^2} + \frac{5}{6}A^2_\zeta\frac{N_{\alpha\beta}N_{\beta\gamma}N_{\gamma\alpha}}{(N_{\alpha}N_{\alpha})^3}\ln(kL)\,.
 \label{eq:fnl}
\en
The second, $k$-dependent term can be thought of as a loop correction and is usually subdominant.

The current observational bounds on the observables at $68\%$ confidence level are~\cite{Ade:2013zuv}
\be 
 10^9\times A_\zeta^2 = 2.1886^{+0.0532}_{-0.0583}\,,
 \label{obsAs}
\en
for the amplitude of the perturbations,
\be
 \ns &=& 0.9603\pm0.0073\,,
 \label{obsns}
 \en
for the spectral index and~\cite{Ade:2013ydc}
\be
 \fnl &=& 2.7\pm5.8\,,
 \label{obsfnl}
 \en
 for the reduced bispectrum.
 
\section{Analytical Estimates}
\label{sec:anes}

In Section~\ref{sec:numerical} we will calculate the observables $A_\zeta$, $\ns$ and $\fnl$ numerically, but in order to develop some physical intuition about the dynamics, it is instructive to consider a simple analytical approximation first.

The total expansion from the initial hypersurface $*$ to the final uniform-density hypersurface $f$ consists of three separate stages. Initially $\chi=0$ and both $\phi<0$ and $\sigma$ slow-roll towards the minimum of their quadratic potential, until the trajectory hits the critical surface region where $\chi$ becomes tachyonic, i.e., $m_\chi^2$ becomes negative. We denote the amount of expansion during this first stage by $N_{\rm{crit}}$. It is followed by a short period of tachyonic growth of $\chi$, which lasts until backreaction makes the dynamics non-linear, and during which the universe expands by $N_{\rm{tach}}$ e-foldings. We assume that after this, the fields equilibrate instantaneously, and the universe expands with the equation of state of radiation to the final hypersurface. This expansion is denoted by $N_{\rm rad}$.
The total amount of expansion is therefore
\begin{equation} 
\begin{split}
 N&(\phi_*,\sigma_*)=  \\
 & N_{\rm crit}(\phi_*,\sigma_*)+N_{\rm tach}(\phi_*,\sigma_*)+N_{\rm rad}(\phi_*,\sigma_*).
\label{n}
\end{split}
\end{equation}

Let us first compute the number of e-folds $N_{\rm{crit}}(\sigma_*)$ from a flat hypersurface at horizon crossing to the critical surface defined by Eq.~(\ref{circle}), i.e. $m^2_\chi=0$.

During this period, the fields are slowly rolling, and 
the equations of motion are therefore
\be \frac{\partial\phi}{\partial N}&=&-\frac{m^2\phi-\alpha mv^2/\sqrt{2}}{3H^2},\nn
    \frac{\partial\sigma}{\partial N}&=&-\frac{m^2\sigma}{3H^2},\label{eom}
\en
where 
\be
\mpl^2 H^2 = \frac{\alpha^2v^4}{12} + \frac{m^2(\phi^2+\sigma^2)}{6} - \frac{\alpha m v^2\phi}{3\sqrt{2}}.
\label{hubble}
\en
Taking the ratio between the equations in Eqs.~(\ref{eom}) and solving for $\sigma$ with initial condition $\sigma(\phi_*)=\sigma_*$, we find 

\be
\sigma(\phi) = \sigma_*\frac{\phi - w}{\phi_*-w}\,.
\label{sigmaphi}
\en
We then substitute Eq.~(\ref{sigmaphi}) into Eq.~(\ref{hubble}) to find

\be
\begin{split}
N&_{\rm crit}(\phi_*,\sigma_*)=\frac{1}{4}\left(1+\frac{\sigma_*^2}{(\phi_*-w)^2}\right)\times\\
&\left(\left(\phi_*-w\right)^2-\left(\phi_{\rm crit}(\phi_*,\sigma_*)-w\right)^2\right)\,,
\end{split}
\label{ncrit}
\en
where
\be
\begin{split}
\phi&_{\rm crit}(\phi_*,\sigma_*)=-\frac{1}{w\left(\left(\phi_*-w\right)^2+\sigma_*^2\right)}\times\\
&\left[\left(\phi^*-w\right)\sqrt{v^2\left(v^2+4w^2\right)\left(\phi_*-w\right)^2-4w^4\sigma_*^2}\right.\\
&+\left. v^2\left(\phi_*-w\right)^2-2 w^2\sigma_*^2\right]\,.
\label{fcrit}
\end{split}
\en
is the value of the field $\phi$ at the critical surface (\ref{circle}).
Thus, Eq.~(\ref{ncrit}) gives the amount of inflation happening before the transition starts.

Next, we consider the expansion $N_{\rm tach}$ during the linear tachyonic growth of $\chi$.
Assuming that $\phi$ and $\sigma$ are homogeneous, 
we can approximate
\be
\phi(t) &\approx& \phi_{\rm{crit}} - \dot{\phi}t.\nn
\sigma(t) &\approx& \sigma_{\rm{crit}} - \dot{\sigma}t.
\label{linear}
\en
The time dependence of the fluctuations $\ctk$, up to a short time after the transition, is well approximated by the Minkowski linearized equation of motion
\begin{equation}
\begin{split}
 \partial_t^2&\ctk =\\
&= \left[\frac{\alpha^2}{2}\left( 2v^2\!-\!\phi^2(t)\!-\!\sigma^2(t)\right)-\sqrt{2}\alpha m\phi(t)-k^2\right]
\ctk \\
                    &\approx  \left[\left((2\alpha^2\phic+\sqrt{2}\alpha m)\dot{\phi}+2\alpha^2\sigmac\dot{\sigma}\right)t-k^2\right]\ctk .
\label{chieom}
\end{split}
\end{equation}
Each mode $k$ starts growing when $m^2_k\equiv\omega^3 t - k^2$ becomes negative, that is at $t_k =k^2/\omega^3$.

In a quantum field theory, Eq.~(\ref{chieom}) is valid as an operator equation. The initial state is the vacuum and at tree level it is completely described by the two-point functions of the fields,
\be \langle \chi^*(\mathbf{k})\chi(\mathbf{k'})\rangle &=& \frac{1}{2|\mathbf{k}|}(2\pi)^3\delta^3(\mathbf{k}-\mathbf{k'}),\nn
    \langle \pi^*(\mathbf{k})\pi(\mathbf{k'})\rangle &=& \frac{|\mathbf{k}|}{2}(2\pi)^3\delta^3(\mathbf{k}-\mathbf{k'}),
    \label{vac}
\en
where $\pi=\delta_t\chi$.

Following Ref.~\cite{tacpreh2} we find that, at late times, the fluctuations of $\chi$ go as 
\be 
\langle \chi^2 \rangle (t) \approx \frac{2.645}{128\pi^3}\frac{\omega}{t}\exp\left[\frac{4}{3}(\omega t)^{3/2}\right],
\label{equ:tachyonic}
\en
where
\begin{equation}\omega = \left[(2\alpha^2\phic+\sqrt{2}\alpha m)\dot{\phi}+2\alpha^2\sigmac\dot{\sigma}\right]^{1/3}.
\end{equation}
We are interested in the time $t_{\rm br}$ at which the linear approximation (\ref{chieom}) fails and the dynamics
become non-linear. We approximate that this happens when $\langle \chi^2 \rangle  = v^2$. 

Using Eq.~(\ref{equ:tachyonic}), the asymptotic late-time solution gives
\be N_{\rm{tach}}&=&Ht_{\rm br}\approx\frac{H}{\omega}\left(\frac{3}{2}\ln\frac{v}{\omega}\right)^{2/3}.
\en
Using Eqs.~(\ref{sigmaphi}) and~(\ref{fcrit}) we can express $H$ as a function of $\phi_*$ and $\sigma_*$. 

Finally, we assume that from $t_{\rm br}$ onwards, the universe is radiation dominated. 
Therefore the amount expansion from $t_{\rm br}$ to the final hypersurface with constant energy density $\rho_f$ is
\begin{equation}
N_{\rm rad}=\frac{1}{4}\ln\frac{\rho_{\rm br}}{\rho_f},
\end{equation}
where $\rho_{\rm br}=\rho(t_{\rm br})$.
From this we find
\be
\frac{\partial N_{\rm rad}}{\partial \sigma_*}=\frac{1}{4}\frac{\partial \log \rho_{\rm br}}{\partial \sigma_*},
\en
where $\rho_{\rm rad}$, analogously to $H$ above, depends on the field values at horizon crossing.

We are now ready to compute the observables of Eqs.~(\ref{eq:ampl},\ref{eq:spectrum},\ref{eq:fnl}). 
The analytical expressions for the observables are long and complicated. However, there exists a limit in which they become simple enough to be presented here. Indeed, for trajectories such that $\sigmac=0$, the derivatives of $N_*^f$ are dominated by the contribution of $N_{\rm crit}(\phi_*,\sigma_*)$ and reduce to

\be 
\frac{\partial N_*^f}{\partial \phi_*} = \frac{\partial N_{\rm crit}}{\partial \phi_*} &=& \frac{1}{2}\left(\phi_* - \frac{v}{\sqrt{2}c}\right)\label{Nphi}\,,\\
\frac{\partial N_*^f}{\partial \sigma_*}\simeq \frac{\partial N_{\rm crit}}{\partial \sigma_*} &=& 0\label{Nsigma}\,.
\en
Eq.~(\ref{Nsigma}) is not exactly zero because of the contribution of $N_{\rm rad}$. However, in practice both this term and the loop correction in Eq.~(\ref{eq:fnl}) are negligible. Therefore the observables can be expressed using Eq.~(\ref{Nphi}) and its derivative with respect to $\phi_*$ as
 
\be
\ns 
    &=& 1 - \frac{16c^2\mpl^2}{(v-\sqrt{2}c\phi_*)^2}\,,\label{ns0}
\en
and
\be
\fnl
     &=& \frac{10}{3}\frac{c^2\mpl^2}{(v-\sqrt{2}c\phi_*)^2} \,.\label{fnl0}
\en
In the limit $c\ll v/\phi_*$, we find the simple relations $\ns= 1-16c^2\mpl^2/v^2$ and $\fnl=10c^2\mpl^2/3v^2$. The opposite regime, when $\phi_*$ dominates the denominator, gives $\ns = 1 - 8\mpl^2/\phi_*^2$ and $\fnl= 5\mpl^2/(3\phi_*^2)$.
Therefore, the presence of a third scalar field which is light at horizon crossing generates a red spectral index in this model, in contrast to the original hybrid inflation scenario. In the limit in which the potential (\ref{vs}) is vacuum dominated at horizon crossing, this is due to the linear term in $\phi$, that implies $\epsilon_*\simeq\eta_*$ and $\ns<1$. As we approach $\phi_*$ domination, the spectral index decreases.

Planck results for the amplitude of the primordial curvature perturbation Eq.~(\ref{obsAs}) constrain the model's parameter space. During the slow-roll phase, the amplitude is the only observable that depends on rescalings of the potential, meaning we can fix it to the observed value \emph{a posteriori}. However, once inside the critical surface Eq.~(\ref{circle}), the velocity of the transition depends on the scale of the potential, implying that observables such as $\fnl$ and $\ns$ are not invariant under potential rescalings. Fortunately, their dependence on $\sigmac$ does not change qualitatively as we vary $\alpha$.

\section{Numerical Results}
\label{sec:numerical}

We have to keep in mind that the analytical results rely on the linear approximation (\ref{linear}) and are therefore at best indicative.
Nevertheless, we can use them to identify suitable parameters for detailed numerical study.

In order to calculate the perturbations numerically, we solved the homogeneous equation of motion for the three fields,
\be
\ddot{\varphi}_\alpha + 3H\dot{\varphi}_\alpha+\frac{\partial V}{\partial \varphi_\alpha} &=& 0\,, 
\label{numeom}
\en
where $\varphi_\alpha = (\phi,\sigma,\chi)$. 
Because we are ignoring inhomogeneous fluctuations, our calculations do not capture non-equilibrium phenomena that take place at the end of inflation, 
 which could lead to interesting effects.
          
In order to solve the equations (\ref{numeom}), we have to choose initial conditions for the fields. 
Because $\chi$ is heavy initially, i.e. $m_\chi>H$, it does not have superhorizon fluctuations. Therefore inflation happens in the $\phi-\sigma$ plane. We choose the initial conditions at horizon crossing $\phi_*$ and $\sigma_*$, and solve the equations until the trajectory oscillates around the SBV. After a few oscillations, we stop the simulation and record the final energy density $\rho_f$. Then, we vary the initial conditions $\phi_*$ and $\sigma_*$ and repeat the simulation, ending it when the energy density equals $\rho_f$. We repeat the process until we have enough points to compute the derivatives of  $N(\phi_*,\sigma_*)$ needed to evaluate the observables $A_\zeta$, $\ns$ and $\fnl$ [Eqs.~(\ref{eq:ampl}),~(\ref{eq:spectrum}) and (\ref{eq:fnl}) respectively]. At this point, we compare the numerical $A_\zeta$ to the Planck constraint (\ref{obsAs}) and rescale $\alpha$ as required. Finally, we repeat the process with the correct $\alpha$, obtaining a data point.

In practice, $\chi$ also has to be given small non-zero initial value because $\chi=0$ is a classically stable trajectory protected by symmetry. Therefore if $\chi=0$ initially, the system never reaches the SBV. In reality, the symmetry breaking takes place because of quantum fluctuations, but in the absence of inhomogeneous fluctuations, Eq.~(\ref{numeom}) fails to capture this. However, since our aim is to simply study the curvature perturbation generated on super-horizon scales by the light scalars $\phi$ and $\sigma$, we only need to know how the time necessary for the phase transition to the SBV depends on the field 
values at horizon crossing $\phi_*$ and $\sigma_*$. The dependence can be extracted by choosing a fixed initial value  $0<\chi_{\rm i}\ll v$ and varying the initial conditions at horizon crossing in the $\phi-\sigma$ plane, giving the function $N(\phi_*,\sigma_*)$.

\begin{figure}
\centering
\includegraphics[width= 8cm]{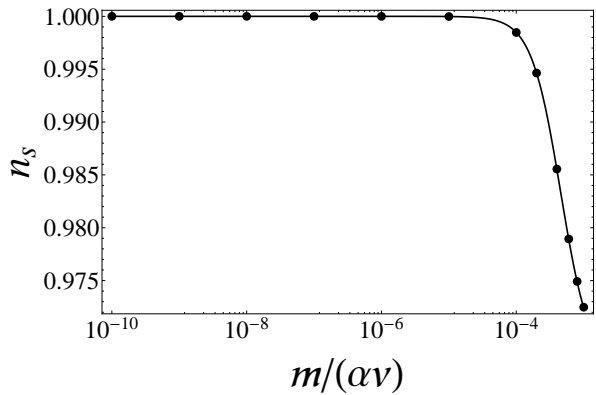}
\caption{$\ns$ as a function of $c\equiv m/(\alpha v)$. The continuous line represents the analytical result  Eq.~(\ref{ns0}) and the dots are numerical data points. The other parameters are fixed as $v=10^{-2}\mpl$, $\sigmac = 10^{-4}\mpl$ and $\alpha$ is chosen to satisfy the Planck amplitude constraint. Although the lowest data point, $\ns\simeq0.972$ and $c=10^{-3}$, is still outside the Planck $68\%$ confidence limit range, the spectral index decreases as the inflaton $\phi$ becomes heavier.}
\label{nsvsc}
\end{figure}

\begin{figure}[t]
\centering
\includegraphics[width= 8cm]{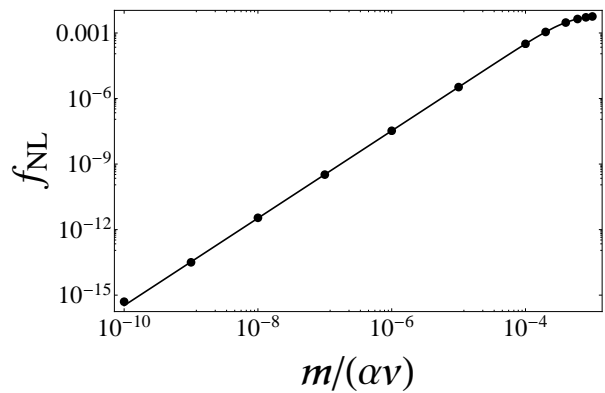}
\caption{$\fnl$ as a function of $c\equiv m/(\alpha v)$. The continuous line represents the analytical result Eq.~(\ref{fnl0}) and the dots are numerical data points. The other parameters are fixed as $v=10^{-2}\mpl$, $\sigmac = 10^{-4}\mpl$ and $\alpha$ is chosen to satisfy the Planck amplitude constraint.}
\label{fnlvsc}
\end{figure}

\begin{figure}
\centering
\includegraphics[width= 8cm]{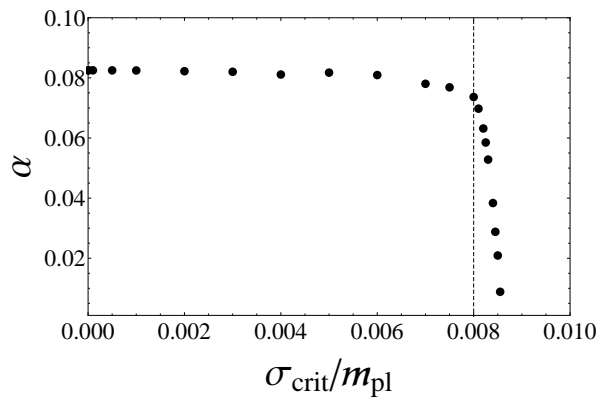}
\caption{$\alpha$ as a function of $\sigmac$ after the rescaling necessary to fit the observed amplitude of perturbations. The other parameters are fixed as  $v=10^{-2}\mpl$ and $m= 10^{-3}\alpha v$. The region to the left of the vertical dashed line leads to non-Gaussianity parameter $\fnl$ within the Planck $68\%$ confidence limit range.}
\label{scalpha3}
\end{figure}

\begin{figure}
\centering
\includegraphics[width= 8cm]{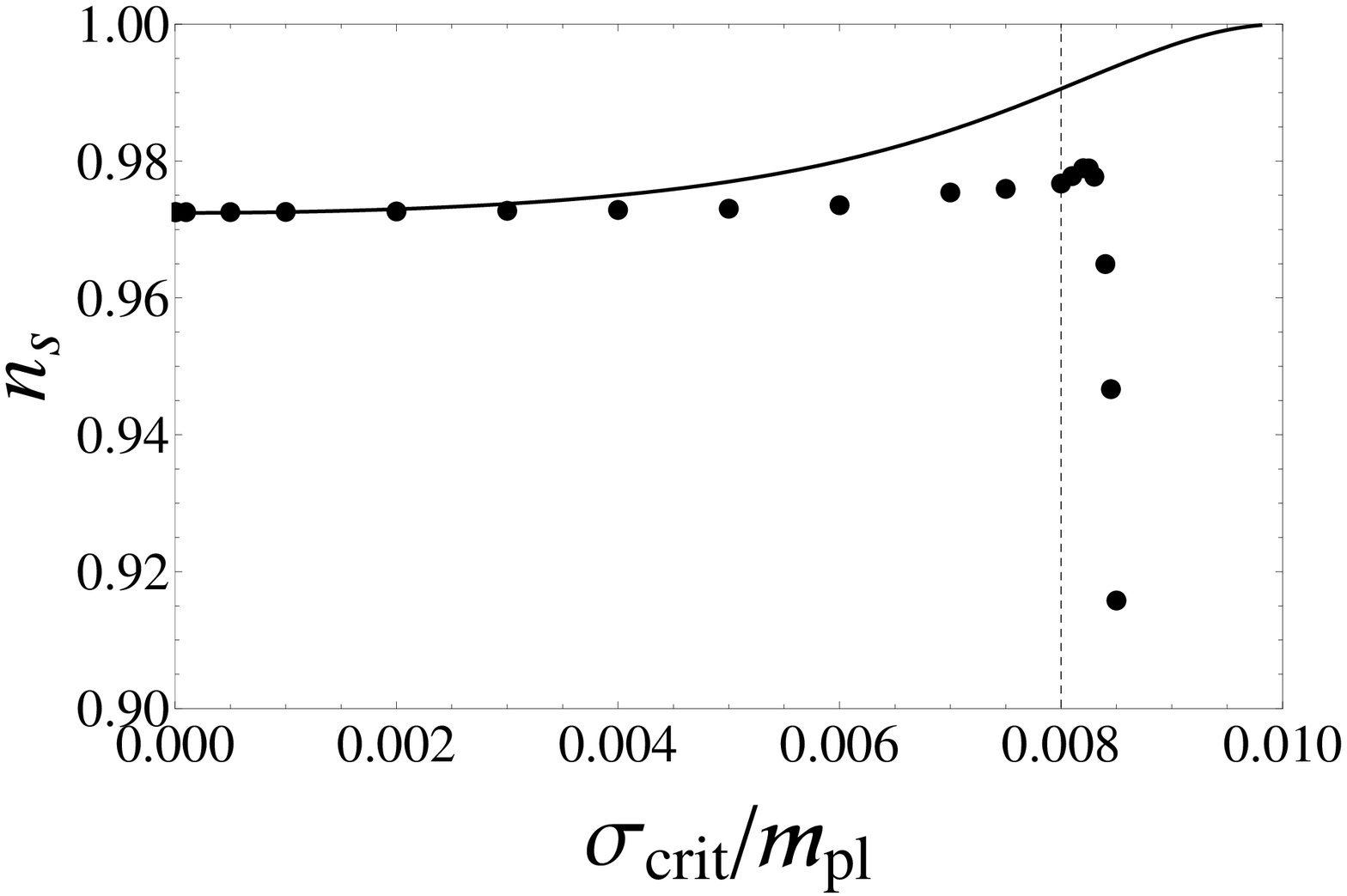}
\caption{$\ns$ as a function of $\sigmac$, numerical results (dashed line) and analytical results (continuous line). The other parameters are fixed as $\alpha=10^{-1}$, $v=10^{-2}\mpl$ and $m= 10^{-3}\alpha v$. The region to the left of the vertical dashed line leads to non-Gaussianity parameter $\fnl$ within the Planck $68\%$ confidence limit range.}
\label{ns3}
\end{figure}

\begin{figure}
\centering
\includegraphics[width= 8cm]{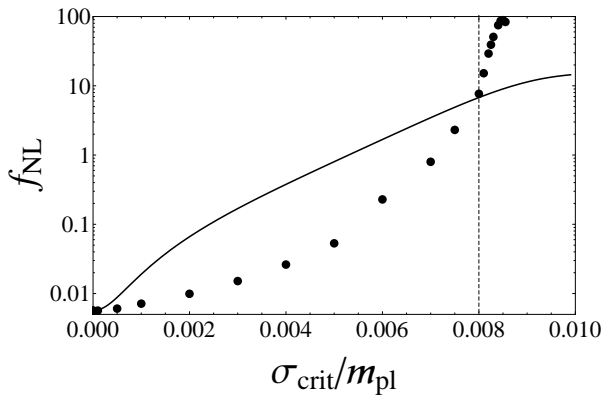}
\caption{$\fnl$ as a function of $\sigmac$, numerical results (dashed line) and analytical results (continuous line). The other parameters are fixed as $\alpha=10^{-1}$, $v=10^{-2}\mpl$ and $m= 10^{-3}\alpha v$. The region to the left of the vertical dashed line leads to non-Gaussianity parameter $\fnl$ within the Planck $68\%$ confidence limit range.}
\label{fnl3}
\end{figure}

The numerical results confirm the validity of the analytic estimates Eqs.~(\ref{ns0}) and (\ref{fnl0}) at small $\sigma_{\rm crit}$, as is shown in Figs.~\ref{nsvsc} and \ref{fnlvsc}. Furtmermore, requiring the spectral index to be within the $95\%$ confidence limit leads to the constraint $c\gtrsim5\times10^{-4}$ for $v=10^{-2}\mpl$. For fixed $c$, considering smaller values of $v$ drives the spectral index closer to unity, as can be seen from Eq.~(\ref{ns0}). On the other hand, larger values lead to smaller $\ns$, increasing the agreement with observations. However, as can be seen from Eq.~(\ref{eom}), increasing $v$ implies that $\phi<0$ rolls faster through the critical surface, decreasing the cross section of field values ending up in the SBV.

Eq.~(\ref{gamma}) highlights that as $c$ increases, so does the likelihood of a random trajectory hitting the critical surface. However, increasing $c$ has another undesired effect: the light scalars $\phi$ and $\sigma$ roll faster compared to $\chi$ and the transition to the SBV might not have time to happen if too close to the edge of the critical surface. Indeed, we found that for $c>10^{-3}$, only trajectories significantly far from the edge hit the SBV.
   
From Fig.~\ref{nsvsc} we know that for $c<10^{-3}$ the predicted spectral index is in conflict with observations (\ref{obsns}) in the limit $\sigma \rightarrow 0$. For $c=10^{-3}$, we find $\ns\simeq0.972$, which is also excluded by the $68\%$ confidence limit bounds, but is well within the $95\%$ bracket. Furthermore, Fig.~\ref{fnlvsc} shows that $\fnl$ grows proportionally to $c$.

Considering these constraints we choose $v=10^{-2}\mpl$ and $c=10^{-3}$ for more detailed study. For such parameters, the loop correction in Eq.~(\ref{eq:fnl}) is of order $10^{-8}$ in the limit $\sigmac\rightarrow 0$.

Fig.~\ref{scalpha3} shows the rescaled dimensionless parameter $\alpha$ as a function of $\sigmac$. It is interesting to note that it is at least of order $10^{-2}$, and of order $10^{-1}$ for trajectories leading to a viable cosmology.  

In Figs.~\ref{ns3} and \ref{fnl3} the spectral index $\ns$ and the non-Gaussianity parameter $\fnl$ are plotted as a function of $\sigmac$. The dots represent numerical data points whereas the continuous line shows the analytical results. We can see that the analytical approximation is only valid for small $\sigmac$, and its accuracy decreases with the increasing importance of non-linear dynamics.
For $c=10^{-3}$, we find that the universe reaches the SBV at the relatively low probability $P_{\rm SBV} \approx 4\times10^{-4}$.

Requiring the spectral index to be in the observationally compatible range at $95\%$ confidence level constrains the viable range of field values to $|\sigmac| \lesssim 8.3\times10^{-3}\mpl$. The non-Gaussianity parameter $\fnl$ increases the constraint on the initial conditions, giving $|\sigmac| \lesssim 8\times10^{-3}\mpl$ for predictions within the $68\%$ confidence limit bracket. 

Note that for $\sigmac>8.3\times10^{-3}\mpl$ the spectral index drops sharply, as can be seen in Fig.~\ref{ns3}. 
The drop is due to the increase in magnitude of $N_{,\sigma\sigma}/N_{,\sigma}$ as $\sigmac$ approaches the edge of the circle (\ref{circle}). Since $\dot{\sigma}_*$ is negative, this generates a negative contribution to the spectral index, as can be seen from Eq.~(\ref{eq:spectrum}).

\section{Conclusions}

In this paper we have studied a SUSY F-term version of hybrid inflation with three dynamically relevant scalar fields: the inflaton $\phi$, the waterfall field $\chi$ and a light scalar field $\sigma$. Softly broken R-symmetry ensures that the scalar $\sigma$ naturally has a light mass.

The potential (\ref{vs}) has three degenerate global minima, one with unbroken and two with broken symmetry, as illustrated in Fig.~\ref{Vphichi}. Inflationary trajectories that end up in the symmetry breaking vacuum require fine tuning of initial conditions, which could be justified by anthropic arguments. For parameter values leading to observationally compatible predictions, we found that, under reasonable assumptions, the likelihood of an inflationary trajectory reheating in the symmetry breaking vacuum is of order $4\times10^{-4}$.

We considered the regime in which the symmetry breaking field $\chi$ is heavy during the inflationary phase, i.e. $m_\chi>H$, and does not affect super-horizon observables. We have shown that the presence of the light scalar $\sigma$ generates a red spectral $\ns<1$, as can be seen in Eq.~(\ref{ns0}), in agreement with observations and in contrast with the original hybrid inflation scenario. 

The analytical analysis carried in Section~\ref{sec:anes} led us to choose parameters $v=10^{-2}\mpl$ and $c\equiv m/(\alpha v)=10^{-3}$ for a detailed study. In Section~\ref{sec:numerical}, we numerically computed the observables $\ns$, $\fnl$ and $A_\zeta$ for such parameters. We found that trajectories hitting the critical surface Eq.~(\ref{circle}) at values such that $|\sigmac|\leq 8\times10^{-3}\mpl$ lead to a spectral within the current $2\sigma$ confidence limit range and to non-Gaussianity parameter $\fnl$ within the $68\%$ confidence limit range, as shown, respectively, in Figs.~\ref{ns3} and \ref{fnl3}. Furthermore, for such trajectories, the rescaled dimensionless parameter $\alpha$ is of order $10^{-1}$, as can be seen in Fig.~\ref{scalpha3}.

It is known that the presence of light scalar fields has non-trivial effects on preheating dynamics~\cite{Bethke:2013aba,preh}. Analogously, light scalars should affect the non-equilibrium dynamics of tachyonic preheating, leading to potentially observational signatures. Their study requires a lattice simulation of the symmetry breaking phase transition, which is beyond the scope of this paper.

\section{Acknowledgements}
AR is supported by the STFC grant ST/J000353/1 and SO by the Swiss National Science Foundation.
This work was also supported by the Royal Society International Joint Project JP100273.

\end{document}